\documentclass[fleqn,12pt,twoside]{letter}
\usepackage{espcrc1}

\usepackage{graphicx}

\newcommand{\be}{\begin{equation}}
\newcommand{\ee}{\end{equation}}
\newcommand{\ba}{\begin{array}}
\newcommand{\ea}{\end{array}}

\newcommand{\bx}{{\bf x}}
\newcommand{\by}{{\bf y}}

\newcommand{\Pq}{\psi}
\newcommand{\Pqb}{\bar{\psi}}

\newcommand{\coes}{{\c c\~oes }}

\title{Di-Antiquarks condensation 
in Color Superconductivity }

\author{ F\'abio L. 
Braghin\address[IF-USP]{Instituto de F\'\i sica, Universidade de
        S\~ao Paulo,\\
        C.P. 66318, CEP 05315-970, S\~ao Paulo, Brazil } \\
(email: braghin@if.usp.br) 
        \thanks{Work supported by FAPESP, SP, Brazil.
} 
}

\begin{document}

\maketitle

\begin{abstract}
Some consequences of a classical vector field 
(chromo-electromagnetic field)
coupled to quarks,  which undergo to superfluid and/or
superconductive states with diquark / diantiquark
 condensation,  are investigated
For this,
one scalar field  exchange is considered 
in the lines investigated by
Pisarski and Rischke \cite{PISARSKI-RISCHKE}
in the mean field approach. 
Some effects and possible consequences  are discussed.
\end{abstract}

%

\section{ Introduction}

In the high energy density regime
of the  phase diagram of matter,
hadrons are not expected to remain as confined bound states 
due to deconfinement and chiral symmetry restoration.
In particular at low temperatures,
several configurations of diquarks (condensates)
 are expected to appear in different channels
(such as scalar and vector), due to attractive
color interaction (in the triplet channel) 
 although color superconductivity might occur
without a GAP \cite{CSCD-98,BUBALLA}.
Nambu-Jona-Lasinio (NJL) and gluon exchange
 models capture many features of 
the dense quark-gluon environment.

In principle diquark condensate would 
differ from the diantiquark condensate 
 for a phase, although 
the distance of antiquarks 
from the quark Fermi surface can
make the di-antiquark to disappear 
\cite{PISARSKI-RISCHKE,ABRIKOSOV-etal}.
Usual calculations do not consider however 
antiquarks, that are produced at high energies, 
which 
can either annihilate or form bound/stationary states 
with quarks \cite{ISMD06+}.
On the other hand from the phase of confined quarks and gluons
to the energies where deconfinement takes place,
gluons can assume an enormous variety of configurations
\cite{gluons}. 
These two aspects  are considered in 
this work.

It is pointed out in this communication that
particular configurations of (classical) vector fields
 \cite{ISMD06+,gluons,SANNINO-SCHAFER}
can produce different
effects including  modifications of the fermion masses,
 chemical potentials 
inducing  diquark/di-antidiquark
condensates at low temperatures.
Results are in agreement with other independent works concerning 
 diquark condensation at low energy densities 
\cite{EBERT-etal}.
The model considers 
the different composite condensates of quarks/antiquarks
are due to a scalar field exchange
(which can represent a phonon) in the mean field approximation.
Possible related effects are discussed 
for Astrophysics and hadrons \cite{IWARA05-RETINHA06-PIC06,ISMD06+}.

\section{ 
Quark condensation 
 due to one scalar boson exchange}

Consider that the exchange of a scalar boson ($\phi$) 
generates  condensates in the framework developed by
 Pisarski and Rischke \cite{PISARSKI-RISCHKE}. With
an additional coupling 
to a classical vector field
the partition function will be given by:
\be \label{Zfunc} \ba{ll}
 Z = N\int {\cal D} \overline{\psi} 
{\cal D} {\psi} {\cal D} {\phi} {\cal D} A_{\mu}
exp \int i \left\{
\overline{\psi}_i
\left( i \gamma_{\mu}{D}^{\mu}_{A}  - \gamma_0 \mu
 + M^*_i \right) \psi_i + {\cal L}_{\phi} 
- \frac{1}{4} F_{\mu \nu} F^{\mu \nu}  \right\},
\ea \ee
where ${\cal L}_{\phi}$ is the Lagrangian density of the massive
scalar field, the effective mass for each quark with label 
$i$ (its quantum numbers)
is
$M^*_i = m_i - g \phi$, a covariant derivative, $D_{\mu}$, can also 
be used for 
the scalar field, if directly coupled to the vector field.
 $\mu$ is the chemical potential
which can be redefined with the  classical component
$\lambda^a A_0^a \equiv A_0$ (where $\lambda_a$ are the GellMann color
matrices) such that the energy eigenvalues and 
equations for each component of quarks/antiquarks
 are differently modified. 
Integrating out the scalar the scalar field 
an effective action is obtained
 which contains fourth order quark-antiquark term.
The shifts of the two-point functions in the 
mean field approach are like:
$\Pq \Pq \to \Pq \Pq - <\Pq \Pq>$ and 
$(\Pqb \Pqb)^{\dagger} \to (\Pqb \Pqb)^{\dagger} -
<\Pqb {\Pqb}>^{\dagger};$
and the same for the $ \psi \Pqb$ functions.
In the limit of $M_s \to \infty$ a NJL model is recovered.
A more complete calculation of the properties 
of the model due to the vector field coupling 
will be shown elsewhere.

The action can be rewritten with doublets of quarks/antiquarks,
using Nambu-Gorkov spinors 
\cite{ABRIKOSOV-etal,PISARSKI-RISCHKE}, 
as:
\be \label{effS} \ba{ll}
\displaystyle{ I = \int 
\frac{d \bx_i}{2}
\left(
\bar{\Psi} A_1(\bx_1, \bx_2) \Psi + 
\bar{\Psi} B_1(\bx_1, \bx_2) \Psi_C + 
\bar{\Psi}_C B_2(\bx_1, \bx_2) \Psi + 
\bar{\Psi}_C A_2(\bx_1, \bx_2) \Psi_C \right)
 ,}
\ea
\ee
where the matrices $A_i,B_i$ are respectively
proportional to the quark-antiquark and diquark/di-antiquark GAPs.
Considering that there is no singularity or topological configuration
a Fourier transformation is performed and
the GAP equations are obtained
\cite{PISARSKI-RISCHKE,JOURNEY2002}. 
In the GAP equations the following
"dressed propagator" emerges:
\be \label{Gff} \ba{ll}
\displaystyle{ (G^{\pm}_{\bar{q}q})^{-1} (k) = 
( (G_0^{\pm})^{-1} - \Delta^{\mp}
G_0^{\mp} \Delta^{\pm} )^{-1} .
}
\ea
\ee
where 
($\Delta^{\pm}$) are
 the diquark/di-antiquark GAPs, the bare
propagator is $G_0^{\pm}$.
The total quark masses are given by: 
$ \tilde{\cal M} = m_q - g_a \int_{p_b} D(p_b-p) <\bar{q} q>^+,$
where
$D(k,{M})$ is the scalar field propagator.
They mix states of chiralities and duplicate the 
number of possible different di-quark/antiquark condensates
Decomposing the GAP into states of  helicity/chirality 
for the quarks/antiquarks,
$\Delta (k) = \sum_{(c=r,l),(h=\pm)}^{(e=\pm)} \phi^{e}_{c,h}$,
 the separated (coupled) equations are obtained.
The corresponding quark/antiquark
dispersion relations are given by:
\begin{equation} \label{energy-dr}
(\epsilon_a^{\pm})^2 [ \phi_{r,+} ] = 
 ( \sqrt{ {\vec{p}}^2 + \tilde{\cal M}^2} \mp 
\mu + g {{A}_0} )^2 
 + \sum_{\alpha}^{i,j} {\cal B}_{\alpha}^{i,j} 
|\phi_{\alpha^{i,j}}|^2,
\end{equation}
where 
there is a combination of each
of the GAPs in the last term. The conditions in which
the component $A_0$ assumes a particular value will not 
be shown elsewhere, although it might be associated to
usual superconductivity \cite{WEINBERG}.
In a first analysis the quark masses (including $<\bar{q}q>$)
are  neglected yielding only four 
possible states of defined chirality.

Results were obtained for massless quarks
when GAP equations are of following form:
\be \ba{ll}
\displaystyle{ \phi_{r,+}^+ = \phi_{l,-}^+ 
(F_0^+(\phi_{l,-})- F_1^+(\phi_{l,-}) )
 + \phi_{l,+}^- (F_0(\phi_{l,+} + F_1(\phi_{l,+})), } 
\ea
\ee
where the remaining integrals in these expressions 
can be solved analytically and they
are of the following types:
\be \ba{ll} \label{integrals}
\displaystyle{
(F_0^{\pm}, \;\;\; F^{\pm}_1) =  
\frac{g^2}{2} \int \frac{d p_0}{2 \pi}
\int \frac{d \vec{p}}{(2 \pi)^3}
\frac{1}{(k-p)^2 - \tilde{M}^2} 
\frac{( 1 \;, \;\;\hat{k}\hat{p} )}{p_0^2 - (\epsilon^{\pm} [\phi ])^2 }
.}
\
\ea
\ee
It  will be considered nearly zero exchanged momentum,
yielding constant $(\Delta^{\pm}, \phi^{\pm})$.

\subsection{  External chromo-electromagnetic fields}

The vector field Euler Lagrange equation can be written as:
$  D_{\mu} F^{\mu \nu}_a  = 
(\Pqb (\bx) \gamma^{\nu}
\Gamma \Pq (\by) )_a
,$
where $\Gamma$ can contain color, flavor and spin, 
and the dynamics of the vector  field,  can be 
strongly modified according to the coupling to quarks.
In particular at very high energy densities (chemical potential)
the behavior of chromo(electro)magnetic fields is also quite relevant
 \cite{BUBALLA,EBERT-etal,RISCHKE}.

In the cases considered by Ebert and collaborators \cite{EBERT-etal}
the chromomagnetic (constant or not) field chosen were
those such as:
(1)  $A_i^a = H x_1 \delta_{\mu,2} \delta{3,a}$ ( an abelian
field); 
(2) 
$A_i^a = \sqrt{H/g'} \delta_{i,a}$ ($i=1,2,3$)
 else $A_i^a=0$, leading to  $H_i^a = \delta_i^a H$
(3) $A_1^1 = A_2^2 = \sqrt{H/g} \delta_{i,a}$
yielding $H_i^a = \delta_{i,a} H$.
It is interesting to calculate, as those authors do,
the corresponding energy spectra (for equal chemical potential).
 
In each of these cases, whenever the diquark condensate is formed
there is a corresponding limit 
in which a di-antiquark condensate might appear.
This is not necessarily when $\mu \to -\mu$ because of the 
(external) vector field.
The energy spectra for quarks
and antiquarks in the chromo(electro)magnetic field 
can  have the same modulus and different signs  for
the free fermion case ($\mu = 0$) in some cases.
This yields the same value for the GAP of 
diquark and antidiquark condensates, unless for a phase.
On the other hand, other contributions, mainly from 
the temporal component of 
vector fields, can modify the relative strength
of the attractive interaction / chemical potential 
at lower energies such
as those shown above (contributing to an effective chemical potential
for each component of quarks/antiquarks)
\cite{ISMD06+,IWARA05-RETINHA06-PIC06}.
However the effective chemical potential is not 
the Lagrange multiplier.
The solutions for the (classical) field are the 
key issues in these cases.


\section{Possible consequences}

Whereas the energy needed to a quark to propagate
in the background of a (scalar) quark-antiquark
should be infinitely large due to the confinement,
 (colored) quarks/antiquarks  propagate
in a color-superconductive state either neutral or not
\cite{ALFORD-SCHMITT}.
There must have a maximum amount
of energy for which this quark/antiquark ($q_i/\bar{q}_i$) 
will not disturb/break 
the condensate in which it propagates,
$E_{max}^{q_i} > \Delta_{j,k}$.
The interaction of a quark with $<\bar{q}\bar{q}>$
can produce a local quark-antiquark annihilation,
in which case instabilities would probably arise.
Otherwise it would give rise to quark-antiquark bound states.
In this processes meson production  could be relevant
if energy density is close to the confinement region. 
These mechanisms can be of relevance for 
the structure of dense stars their structure (if
quarks and antiquarks configurations can
coexist or not) and dynamics including energy realise and 
cooling \cite{BUBALLA}.
They will be investigated deeper elsewhere.

Finally, it can be considered the formation of 
di-antiquark codensates in finite size systems
completely akin to 
has been found for diquark condensation \cite{AMORE-etal} 
 associated to a boxes of length of nearly 3 fm. 
This size is almost small enough as 
to permit asking whether
 the different gluon configurations/degrees
of freedom can trigger diquark/di-antiquark condensation
inside hadrons, whose typical volume scale can of the order of 
$4$ fm$^3$ (for a "mean diameter" of a nucleon of the order $d \simeq 2$ fm).
Consequences of hadron production in (intermediary and) high 
energies (low temperatures) 
could be considerable.
Antiquarks could appear
when di-antiquarks condensates are broken, 
mainly as the energy density
approaches the (deconfinement/chiral restoration) 
phase transition point. 
Lattice calculations can
provide a valuable framework for investigating
eventual condensation inside hadrons.
Cosmological consequences would also arise
if one considers hidden antimatter inside hadrons \cite{ISMD06+}.

\noindent {\bf Acknowledgements}

This work   has been 
initiated in the Theory Group (Phys. Dept.) of BNL.
It is based 
in \cite{JOURNEY2002}. 
F.L.B thanks the hospitality of the 
Nuclear Theory Group (BNL)
  and discussions with E. Shuryak, R.D. Pisarski,
D.H. Rischke, C.L. Lima and F.S. Navarra.

\end{document}